\documentclass[letter]{emulateapj}

\usepackage{natbib}

\newcommand{\uKamin}{$\mu$K$_{CMB}$-arcmin}

\begin{document}

\title{Bolocam Observations of Two Unconfirmed Galaxy Cluster
  Candidates from the Planck Early 
  SZ Sample}

\author{J.~Sayers\altaffilmark{1,2}, 
   N.~G.~Czakon\altaffilmark{1},
   C.~Bridge\altaffilmark{1},
   S.~R.~Golwala\altaffilmark{1},
   P.~M.~Koch\altaffilmark{3},
   K.-Y.~Lin\altaffilmark{3},
   S.~M.~Molnar\altaffilmark{4},
   and K.~Umetsu\altaffilmark{3}
 }
\altaffiltext{1}
  {Division of Physics, Math, and Astronomy, California Institute of Technology, Pasadena, CA 91125}
\altaffiltext{2}
  {jack@caltech.edu}
\altaffiltext{3}
  {Institute of Astronomy and Astrophysics, Academia Sinica, 
    P.O. Box 23-141, Taipei 10617, Taiwan}
\altaffiltext{4}
  {LeCosPA Center, National Taiwan University, 
    Taipei 10617, Taiwan}

\begin{abstract}

We present Bolocam observations of two galaxy cluster candidates reported
as unconfirmed in the
Planck early Sunyaev-Zel'dovich (eSZ) sample, PLCKESZ G115.71+17.52 and
PLCKESZ G189.84-37.24.
We observed each of these candidates 
with
Bolocam at 140~GHz from the Caltech Submm Observatory
in October 2011.
The resulting images have white noise levels of 
$\simeq 30$~\uKamin~in their central regions.
We find a significant SZ decrement towards 
PLCKESZ G115.71.
This decrement has a false detection probability
of $5.3 \times 10^{-5}$, and we therefore
confirm PLCKESZ G115.71 as a cluster.
The maximum SZ decrement towards PLCKESZ G189.84 
corresponds to a false detection probability of 0.027,
and it therefore remains as an unconfirmed cluster candidate.
In order to make our SZ-derived results more robust,
we have also analyzed data from the 
Wide-field Infrared Survey
Explorer (WISE) at
the location of each cluster candidate.
We find an overdensity of WISE sources consistent with other
clusters in the eSZ at the location of PLCKESZ G115.71,
providing further evidence that it is a cluster.
We do not find a significant overdensity of WISE sources 
at the location of PLCKESZ G189.84.

\end{abstract}

\section{Introduction}

The thermal Sunyaev-Zel'dovich (SZ) effect describes the
scattering of CMB photons with the hot electrons
in the intra-cluster medium (ICM) \citep{sunyaev72}.
The surface brightness of the SZ effect is independent
of redshift, and the integrated SZ signal
is approximately equal to the total thermal energy
of the ICM divided by the square of the angular diameter distance, 
making the SZ signal 
an excellent cluster mass proxy with little redshift
evolution~\citep{barbosa96, holder00}.
Consequently, surveys using the SZ
effect to detect clusters have an almost mass-limited selection
function and can yield powerful constraints on cosmological
parameters \citep{haiman01,holder01}.  
Three such SZ surveys are currently ongoing, 
conducted by the South Pole Telescope (SPT),
the Atacama Cosmology Telescope (ACT), 
and the Planck satellite,
and each has published initial cluster catalogues
\citep{vanderlinde10, williamson11, marriage11, planck_esz}

The Planck survey has released a list
of 189 cluster candidates called the early SZ (eSZ) sample
\citep{planck_esz}.
Among this list, 169 are previously known clusters,
and 12 were confirmed by the Planck team prior 
to the publication of the eSZ.
Subsequent SZ observations by SPT and the 
Arcminute Microkelvin Imager (AMI) have 
confirmed 5 and 1 additional eSZ candidates,
respectively \citep{story11, hurley-walker11}.
Prior to our original submission of this manuscript,
there had been no published confirmations
of the remaining 2 cluster candidates in the eSZ sample,
PLCKESZ G115.71+17.52 and PLCKESZ G189.84-37.24.\footnote{
AMI observed, but did not detect, 
PLCKESZ G115.71 \citep{hurley-walker11}.
Furthermore, \citet{muchovej12} published an SZA detection of PLCKESZ G115.71,
and an SZA non-detection of PLCKESZ G189.84, after
our original submission of this manuscript.}
We note that \citet{planck_esz} lists possible contamination
from dust emission near PLCKESZ G115.71 and a high level of contamination
from Galactic emission towards G189.84.
Additionally, \citet{planck_esz} specifically notes that
G189.84 is a low reliability source, and is the only
unconfirmed eSZ candidate without a ROSAT All-Sky Survey (RASS) association.

\section{Observations}

\begin{deluxetable*}{ccccccc} 
  \tablewidth{0pt}
  \tablecaption{Bolocam SZ Observation details}
  \tablehead{\colhead{name} & \colhead{noise RMS} &
    \colhead{S/N} 
& \colhead{RA} & \colhead{$\Delta$RA} & \colhead{dec} & \colhead{$\Delta$dec} \\
    \colhead{PLCKESZ} & \colhead{\uKamin} & \colhead{$\zeta$} & 
    \colhead{J2000} & \colhead{arcsec} & \colhead{J2000} & \colhead{arcsec}} 
  \startdata
G115.71 & 29.5 & $5.7$ & 22:26:27.3 & \phn+0.1 & +78:19:15 & +29.7 \\
G189.84 & 36.5 & $3.8$ & 03:59:46.5 & +10.4 & +00:06:34 & +25.8
  \enddata
  \tablecomments{
  From left to right the columns give the name,
  the Bolocam noise RMS, the peak S/N $\zeta$,
  the Bolocam RA/dec centroid for that peak,
  and corresponding offset of this peak
  from the Planck coordinates.}
  \label{tab:one}
\end{deluxetable*}

Bolocam is a mm-wave imaging camera that operates
from the Caltech Submillimeter Observatory (CSO)
with 144 bolometric detectors covering a
circular 8~arcmin field of view (FOV)
with $\simeq 1$~arcmin resolution
at the SZ-emission-weighted band center
of 140~GHz used for these observations~\citep{glenn98, haig04}.
The data described in this manuscript were obtained
by scanning the telescope
in a Lissajous pattern to yield maps with tapered
coverage out to $\simeq 10$~arcmin in radius
centered on the Planck eSZ coordinates
(identical to the observations described in
\citet{sayers11}).
The observations were conducted in October 2011,
almost entirely in very poor weather conditions
with a 225~GHz opacity of $\tau_{225} \simeq 0.20$.
The total integration time was 12 hours for 
PLCKESZ G115.71 and 7.5 hours for PLCKESZ G189.84,
yielding images with central noise RMSs of
29.5~\uKamin~and 36.5~\uKamin.
Note that SPT reached RMSs of $28-39$~\uKamin~in
their targeted
observations of eSZ clusters, which resulted
in detections with peak S/Ns ranging from
6.3 to 13.8 \citep{story11}.

\section{SZ Data Analysis}
\label{sec:SZ}

\begin{figure*}
\centering
\includegraphics[width=.75\textwidth]{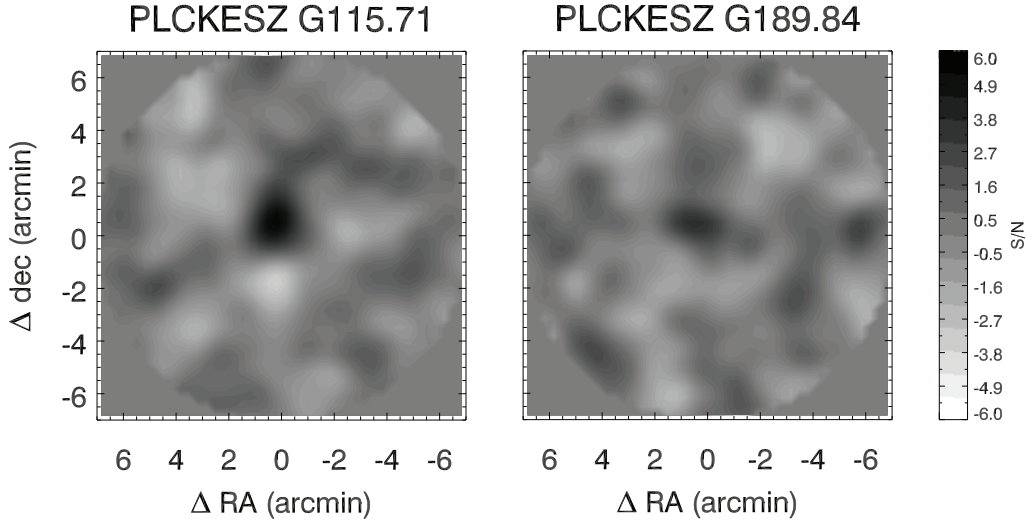}
\caption{
Signal-to-noise thumbnails of the SZ data centered on the Planck
eSZ coordinates. Each thumbnail has been filtered
with the filter that produces the maximum peak S/N $\zeta$
(equal to 5.7 for PLCKESZ G115.71 and 
equal to 3.8 for PLCKESZ G189.84).}
\label{fig:one}
\end{figure*}

These data were reduced according to the procedures
described in detail in \citet{sayers11}, with
only minor changes.
We briefly summarize the data reduction here.
First, we used frequent observations of bright 
quasars to obtain pointing corrections
accurate to 5~arcsec, and we used
observations of Uranus and Neptune to
obtain flux calibration accurate to 5\%.
In particular, the flux calibration has 
been updated based on recent WMAP results
as described in detail in \citet{sayers12}.
The FOV-average signal is then
subtracted at each time sample and the 
data timestreams are high-pass filtered at
250~mHz in order to remove 
atmospheric brightness fluctuations.
Consequently, the cluster images are high-pass
filtered in a way which we 
characterize via simulation to obtain
a signal transfer function for each cluster.

We searched for clusters in these images using
a matched filter according to the formalism
described in \citet{vanderlinde10}
(see also \citet{haehnelt96, herranz02a, herranz02b,
melin06, marriage11}).
We approximate the cluster SZ surface brightness profile $S(\theta)$ as
\begin{displaymath}
S(\theta) \propto (1 + \theta^2/\theta_c^2)^{-1},
\end{displaymath}
where $\theta$ is an angular distance and
$\theta_c$ is the core radius.
We then form a filter according to
\begin{displaymath}
\psi(\theta) = \mathcal{FT}^{-1} \left( \frac{\tilde{W}(u) 
\tilde{S}(u) \tilde{B}(u)}{\tilde{N}(u)} \right),
\end{displaymath}
where $\tilde{W}(u)$ is the signal transfer function in
Fourier space, 
$\tilde{S}(u)$ is the cluster profile in Fourier space,
$\tilde{B}(u)$ is the point-spread function in Fourier space,
$\tilde{N}(u)$ is the map-noise power spectrum,
$u$ is an angular frequency,
and $\mathcal{FT}^{-1}$ denotes the Fourier transform back
to map space.
We filter our images in map-space by convolving them
with $\psi(\theta)$, after truncating $\psi(\theta)$
at a radius of 4~arcmin.
This truncation is required to prevent the filter from
introducing noise in the central region of the images
from the low-coverage outer regions.

Analogous to \citet{vanderlinde10},
each image is filtered using 12 different filters,
with $\theta_c$ varied from 0.25 to 3.00~arcmin
in 0.25~arcmin increments (note that our PSF has
a FWHM of $\simeq 1$~arcmin),
and we define our detection 
significance $\zeta$ as the peak S/N
found in any of these filtered images within
an aperture of 4~arcmin radius centered on the
Planck eSZ coordinates.
This search radius was chosen because it corresponds
to the upper limit on the Planck astrometry
for all but the lowest redshift clusters \citep{planck_esz}.
We find $\zeta = 5.7$ for PLCKESZ G115.71
(with $\theta_c = 0.25$) and 
$\zeta = 3.8$ for PLCKESZ G189.84 (with $\theta_c = 0.25$).
The centroids of these maxima are given in 
Table~\ref{tab:one}, and each is within
1~arcmin of the coordinates given in the Planck eSZ.
Figure~\ref{fig:one} shows thumbnails of each
target filtered with the filter that produces
the maximum S/N.

\begin{figure}
\centering
\includegraphics[width=.8\columnwidth]{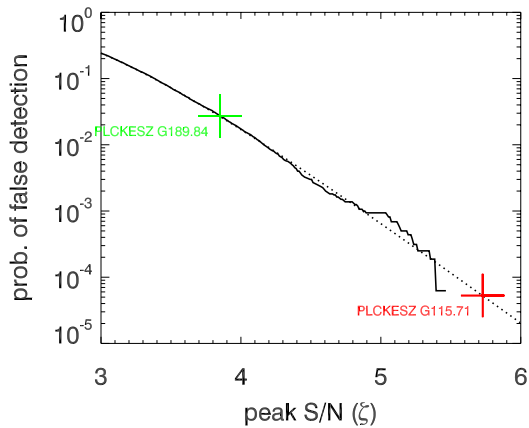}
\caption{The probability of a false detection
within our 4~arcmin search radius as a function of
peak S/N $\zeta$.
The solid line shows the results from simulations
using 16000 noise realizations and the dotted
line is the best fit model to these results.
The values of $\zeta$ for each cluster candidate
are shown as labeled crosses.
Note that, compared to the similar plot 
shown in Fig. 2 of \citet{vanderlinde10}, our
false detection probability at a given value
of $\zeta$ is approximately 3 times larger
compared to SPT.
The likely causes for this difference are
described in Section~\ref{sec:SZ}.}
\label{fig:two}
\end{figure}

We characterize our detection
significance via simulation by generating
noise maps from the combination of 
jackknife realizations of our data
and a model for the astronomical
fluctuations due to the 
cosmic microwave background 
and point sources based on SPT
measurements of the mm-wave power spectrum
\citep{sayers11, reichardt11}.
We generated 1000 such realizations for each
target, along with the 14 other targets observed
with Bolocam in October 2011 to similar depths.
We then searched for the peak S/N in each
of these noise realizations, the distributions
of which were statistically identical
for all 16 targets.
We therefore used all 16000 measurements
to characterize the probability of
measuring a given peak S/N, with the 
results shown in Figure~\ref{fig:two}.
We find the probability of a false detection
of PLCKESZ G115.71 is $5.3 \times 10^{-5}$,
and we therefore
confirm it as a cluster.
Note that PLCKESZ G115.71 has $\zeta = 5.7$,
which exceeds the value of $\zeta$ obtained
in all 16000 of our noise realizations,
we have therefore extrapolated our simulation results
using a parametric fit.
The probability of a false detection 
of PLCKESZ G189.84 is
0.027, and we therefore do not claim
a detection of this candidate.

At a given value of $\zeta$, our false
detection rate per given map area is approximately
three times higher than the SPT false detection
rate described in \citet{vanderlinde10}.
The false detection probability is similar
when $\zeta(\textrm{Bolocam}) \simeq 
\zeta(\textrm{SPT}) + 0.3$.
Although SPT survey data are similar to our
Bolocam data in many respects, there are 
subtle differences that 
contribute to this increase in our false detection
probability.
In particular,
the Bolocam images have tapered coverage and
consequently non-uniform noise properties.
This means that for a given map pixel our
filter will introduce noise from both
lower- and higher-coverage
regions of the map,
and the filter is therefore not as optimal
as the filter used by SPT.

\section{WISE Data Analysis}

\begin{figure*}
\centering
\includegraphics[width=.325\textwidth]{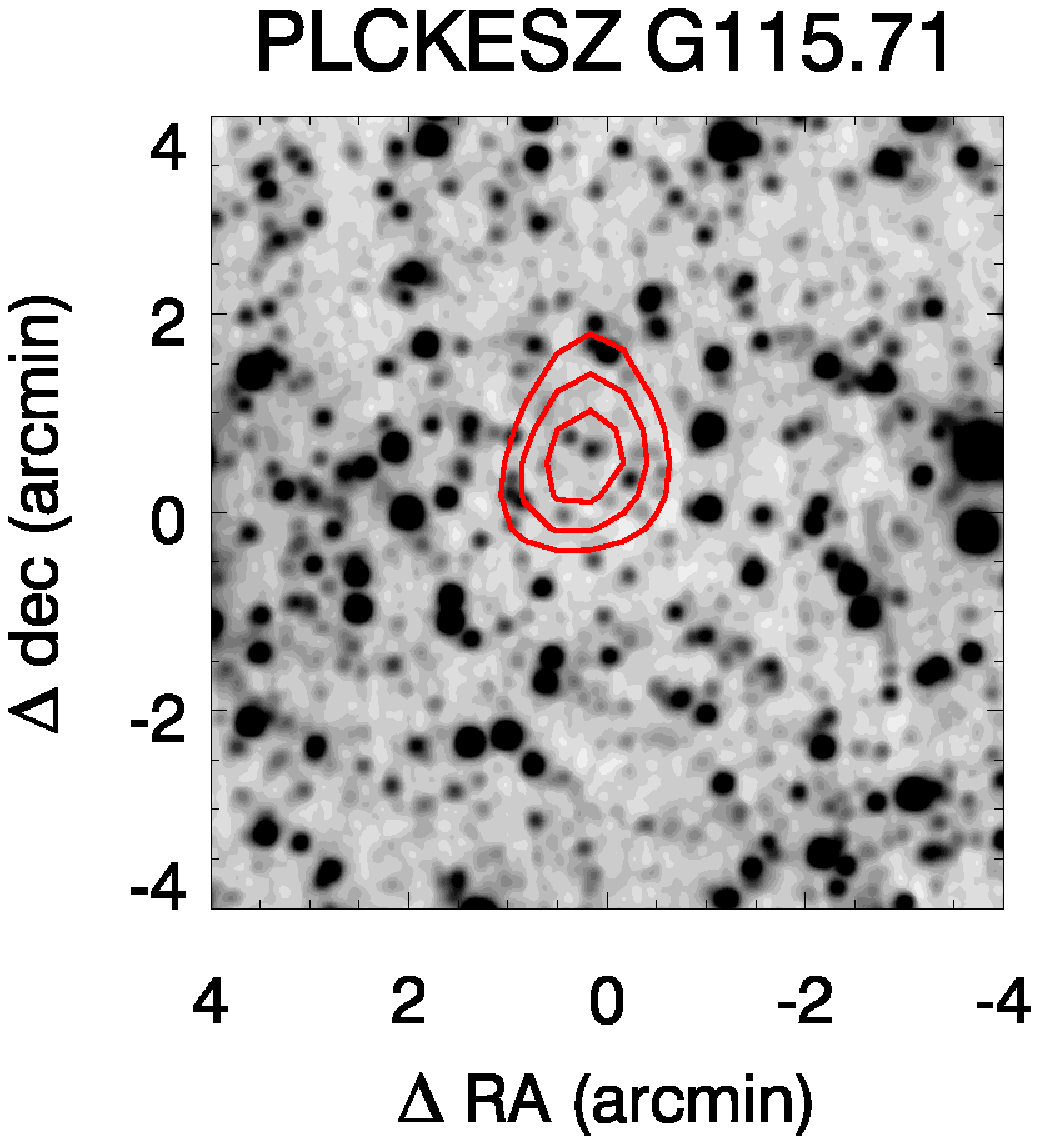}
\hspace{.1\columnwidth}
\includegraphics[width=.325\textwidth]{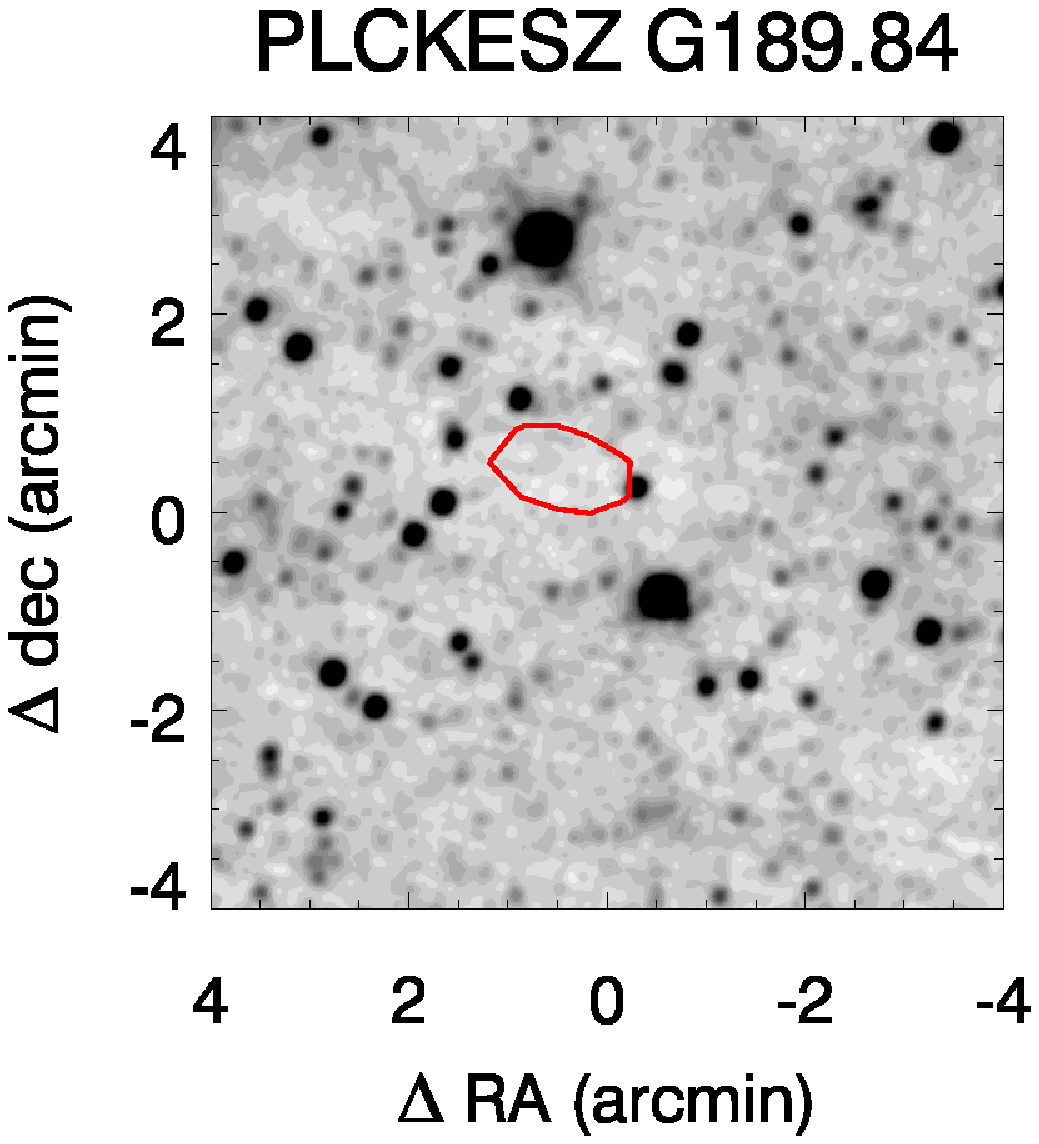}
\caption{WISE band W1 thumbnails centered on the Planck eSZ coordinates
of each eSZ candidate. Overlaid in red is the Bolocam SZ S/N with
a contour spacing of 1 starting with a S/N of $-3$.}
\label{fig:three}
\end{figure*}

\begin{figure*}
\centering
\includegraphics[width=.4\textwidth]{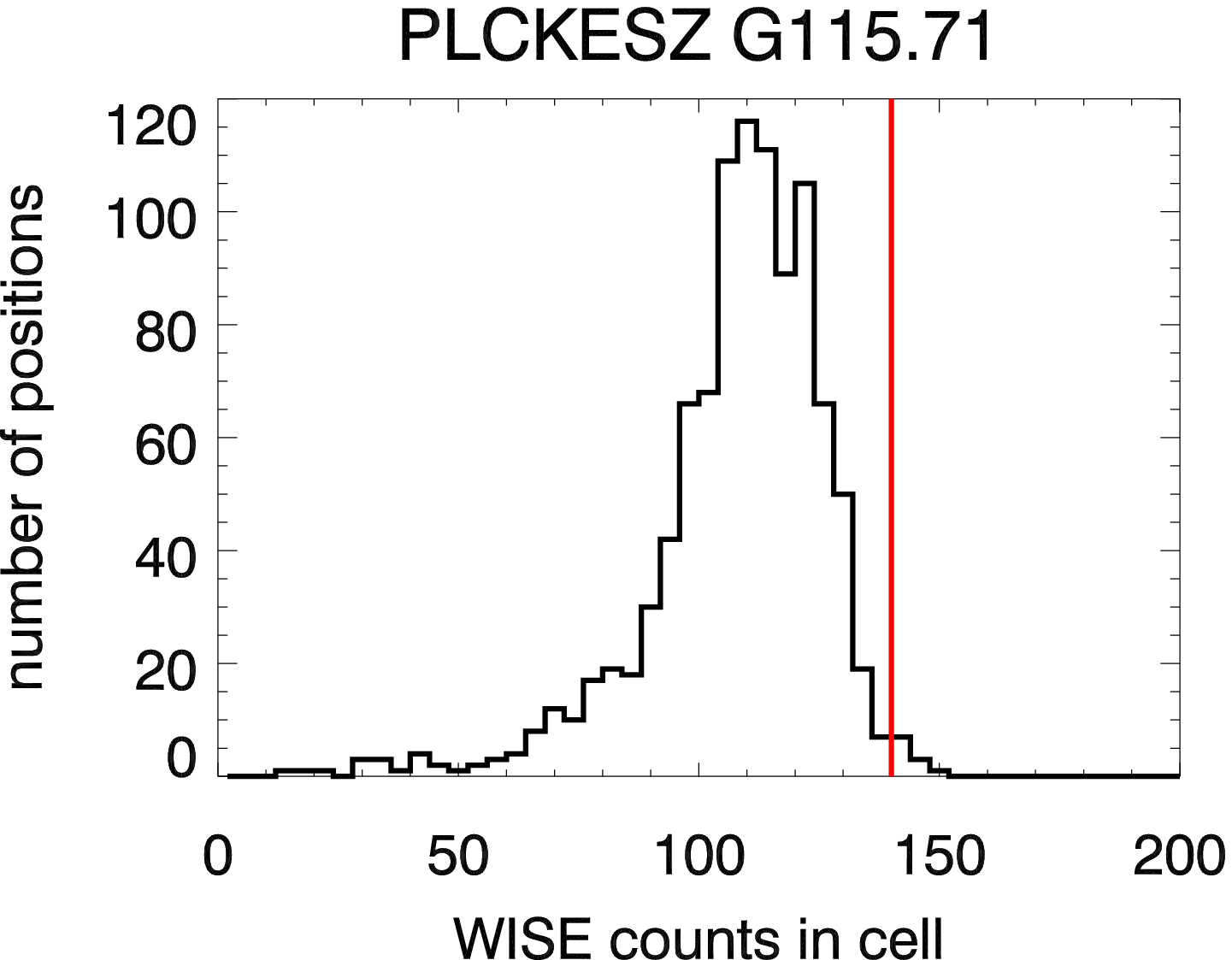}
\hspace{.1\columnwidth}
\includegraphics[width=.4\textwidth]{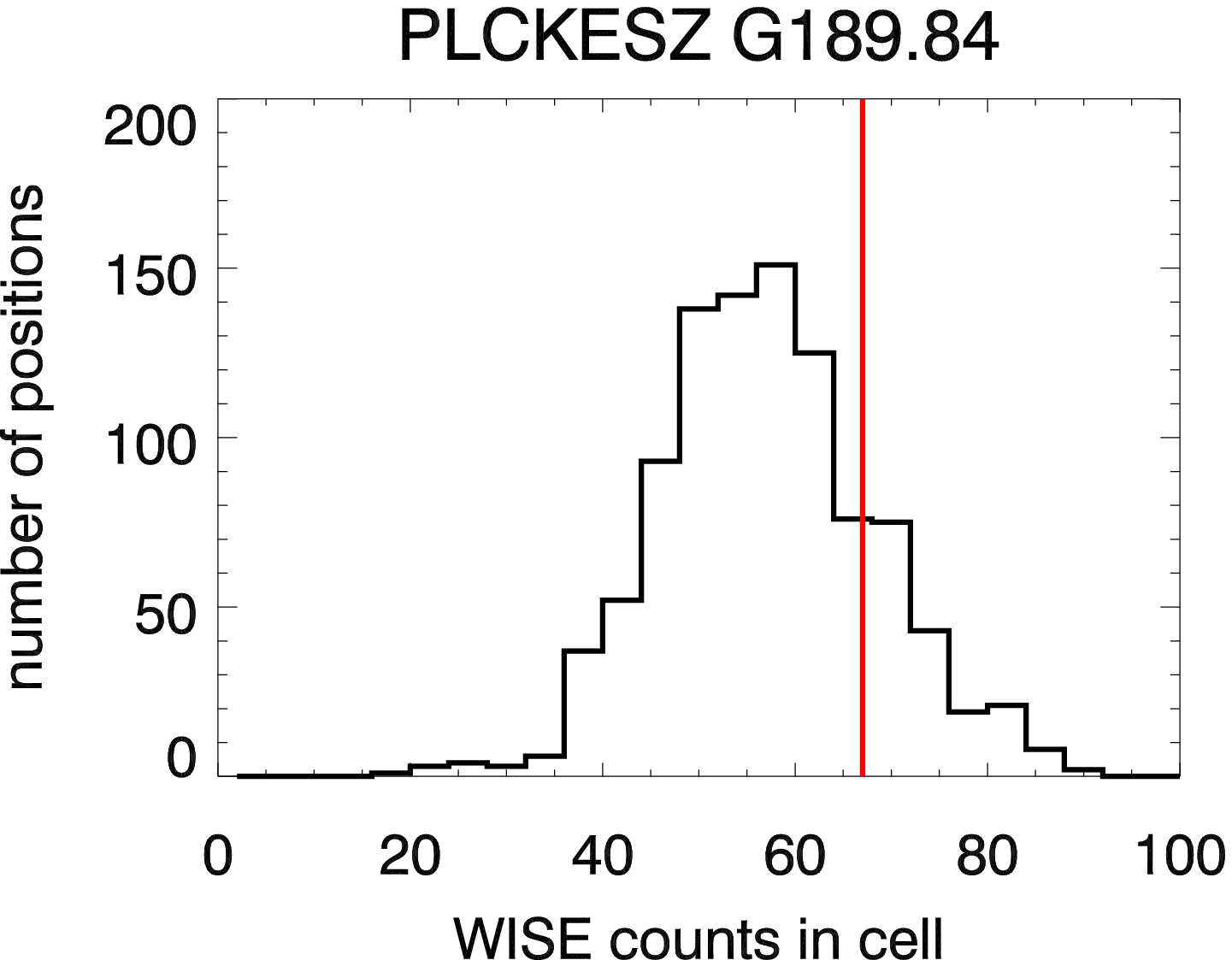}
\caption{The number of WISE sources in the public-release
W1 catalogue within a 3~arcmin radius aperture.
The red line shows the counts centered on the Bolocam
centroid position and the histogram shows the results
for 1000 random positions nearby each eSZ candidate.
Note that the extended tail to the left of each distribution
is due to the WISE bright source masking algorithm.}
\label{fig:wise}
\end{figure*}

We have also analyzed the Wide-field Infrared Survey
Explorer (WISE) public-release data 
\citep{wright10} at the location
of each eSZ candidate.
The central wavelengths of the WISE bands, which are
at 3.4, 4.6, 12, and 22~$\mu$m, are too long to search for these
candidates using a red sequence technique \citep{gladders00},
and so we instead looked for overdensities of sources
in the WISE images.
Specifically, we counted the total number of sources
in the public-release WISE catalogues
for band W1 (3.4 $\mu$m) within a 3~arcmin radius
aperture centered on our Bolocam SZ centroid.
This aperture size is motivated by the fact that it corresponds
to $\simeq 1$~Mpc at $z=0.4$,
the median
redshift of the six clusters originally published
as unconfirmed in the eSZ \citep{story11, hurley-walker11}.
These source counts were then compared to counts 
within 1000 randomly-located and identically-sized 
apertures in the
region within $\simeq 2$~deg of each eSZ candidate.
We limit these blank-sky apertures
to the region immediately surrounding the eSZ candidates
because there are variations in WISE coverage and
in the density of nearby sources,
which causes the blank-sky source density
to vary over the WISE survey.
We find 140 sources within 3~arcmin of PLCKESZ G115.71
compared to a local-blank-sky-average of 108 sources,
and 9/1000 nearby blank-sky regions have more than 140 sources.
We find 67 sources within 3~arcmin of PLCKESZ G189.84
compared to a local-blank-sky-average of 57 sources,
and 168/1000 nearby blank-sky regions have more than 67 sources
(see Figures~\ref{fig:three} and \ref{fig:wise}).

\section{Discussion}

In addition to the two eSZ candidates described in
this manuscript, we have also observed 12 other
eSZ clusters as part of separate Bolocam programs
(Abell 209, Abell 697, Abell 963, Abell 2204, 
Abell 2219, Abell 2261, Abell S1063, MACS J0417.5, 
MACS J0717.5, MACS J1149.5, MACS J1206.2, and
MACS J2211.7).
Results for Abell 697 have already been published in
\citet{sayers11}, and results for the other clusters
will appear in future publications, including a detailed
comparison of the Bolocam and Planck measurements.
The Planck eSZ candidates were selected based on the
mm/submm-wave SZ signal strength measured by Planck,
and the full eSZ catalogue spans a relatively narrow
range of SZ signal strengths.\footnote{
$\simeq 2/3$ of the eSZ candidates have Planck S/N between 
6 and 10 and $\simeq 95$\% of the eSZ candidates have
Planck S/N between 6 and 15.
In particular, the 12 previously known eSZ clusters observed
with Bolocam have Planck S/Ns of 6.8 to 17.4.}
Since Bolocam is sensitive to the mm-wave SZ signal from
these clusters, we therefore expect a correspondingly narrow
spread in the SZ signals measured by Bolocam.
However, we note that the vastly different angular
scales probed by Planck ($\theta \ge 5$~arcmin) and 
Bolocam ($1 \le \theta \le 15$~arcmin) may weaken
the correlation between SZ measurements from Planck and Bolocam.

Since the Bolocam observations of these
previously known eSZ clusters span a range of noise RMSs,
from $16-49$~\uKamin, with peak S/Ns $\zeta = 8.3-21.9$,
we scale the Bolocam values of $\zeta$ to the expected 
value of $\zeta$ at a noise RMS of 
30~\uKamin~($\zeta_{30}$) in order to estimate the absolute
SZ signal measured by Bolocam.\footnote{
As explained in \citet{williamson11},
the SPT peak S/N does not scale linearly
with noise RMS due to changes in the map noise spectrum
and optimal filter shape resulting
from changes in the effective noise ratios of the primary CMB
and the instrument/atmospheric noise.
In contrast, the more severe high-pass filtering
applied to our data significantly reduces the
amount of primary CMB noise in our images,
and our map noise spectrum and 
filter shape are approximately independent
of noise RMS.}
As expected, we find that the values of $\zeta_{30}$ have a relatively
narrow spread, ranging from 10.3 to 20.2
with a median value of $\simeq 14$
(with the exception of Abell 2261, which has $\zeta_{30} = 5.4$).
Consequently, our detection of PLCKESZ G115.71, with
$\zeta = 5.7$ and a noise RMS of 29.5~\uKamin,
indicates that the SZ signal from this cluster
is $\simeq 2.5$ times weaker than the median
SZ signal from Bolocam observations of other
eSZ clusters (although not quite as weak
as the signal from Abell 2261).
Similarly, our image of PLCKESZ G189.84, which
does not contain a significant detection and
has a value of $\zeta=3.8$ and a noise RMS
of 36.5~\uKamin,
indicates that any SZ signal from that location
is at least a factor of $\simeq 3$ times weaker
than the median eSZ cluster. We discuss our
results for each candidate in more detail below.

\subsection{PLCKESZ115.71}

One possible explanation for the weaker-than-expected
SZ signal from PLCKESZ G115.71 is the existence
of positive flux immediately
South of the eSZ coordinates,
which appears to be unresolved
(see Figure~\ref{fig:one}).
This unresolved positive flux has a S/N $\simeq 4$.
We note that there is a source in the 1.4~GHz
NVSS catalogue at 22:26:37.88, +78:17:17.6 (J2000),
which closely matches the location of the positive
flux in our image \citep{condon98}.
However, two nearby sources detected by AMI
at 15~GHz have flux densities of $\simeq 1$~mJy,
indicating that the radio component of
the NVSS source has a sharply
falling spectrum and should not be detected
in our 140~GHz image 
\citep{hurley-walker11}.
In addition, the Planck collaboration
notes possible contamination from dust
emission for this candidate \citep{planck_esz}.
However, due to the low detection significance
of this positive flux in our image, and our lack
of multi-band data, 
it is not clear if this positive
flux is due to astronomical sources, noise,
or both. 
If this positive flux in our image 
of PLCKESZ G115.71
is due to astronomical sources,
then our estimate of both the centroid
and the strength of the SZ signal from
PLCKESZ G115.71 may be biased,
and it may explain why we measure a weaker
than expected SZ signal.

Of the 12 additional eSZ clusters observed by Bolocam,
6 are contained in the currently-available public
WISE data (Abell 2204, Abell 2219, Abell 2261, Abell 697,
MACS J0417.5, and MACS J0717.5).
We counted the number of W1 sources within a 3~arcmin 
radius of each
of these clusters, and again compared those results
to the counts in 1000 nearby background regions.
We find that the median number of 
nearby regions with more WISE sources
than the cluster locations is 6/1000 for these clusters.
Our results for PLCKESZ G115.71 (with more WISE sources than
9/1000 nearby regions) are therefore consistent with
WISE data for previously known eSZ clusters.

\subsection{PLCKESZ G189.84}

Our Bolocam observations towards PLCKESZ G189.84 do
not contain a conclusive detection of an SZ signal;
while there are indications of a decrement, the false
detection probability is 0.027, which is too large
to claim a detection.
Additionally, while WISE data show a slight overdensity
of objects in the direction of this candidate, the
overdensity is consistent with a statistical 
fluctuation (17\% probability to exceed).
Given the lack of a conclusive detection at a signal
level similar to other sources in the eSZ catalogue
there are (at least) three possible explanations
for the observed results:
1) There is a cluster with properties that are unique
compared to others in the eSZ catalogue (weak SZ signal,
weak WISE signal, and no RASS association).
2) The Planck detection, along with the Bolocam and WISE
excursions, are statistical fluctuations, though this
seems unlikely given the expected purity of the eSZ 
catalogue combined with the fact that both the 
Bolocam and WISE excursions are in the direction
of consistency with the presence of a cluster.
3) mm/submm-wave astrophysical contaminants
produced the Planck detection
and caused the non-significant excursion seen in the 
Bolocam data;
this appears to be a reasonable explanation, especially
given the high level of galactic contamination noted by
\citet{planck_esz}.
A definitive conclusion about PLCKESZ G189.84 will
require additional data, in particular results from
the full-depth Planck survey (along with the inclusion
of data from the Planck LFI).

\section{Conclusions}

In October 2011 we performed dedicated
140~GHz Bolocam observations
of two cluster candidates from the Planck eSZ sample,
PLCKESZ G115.71 and PLCKESZ G189.84.
These observations total $\simeq 10$~hours each,
and produce images with a noise RMSs of
$\simeq 30$~\uKamin.
From these images we have confirmed PLCKESZ G115.71
as a cluster at high significance, with a false detection
probability of $5.3 \times 10^{-5}$.
We are not able to confirm PLCKESZ G189.84
as a cluster, although our image does
show a weak decrement with a 
false detection probability of 0.027.
Both our detection of PLCKESZ G115.71, and our
non-detection of PLCKESZ G189.84, indicate
SZ signals significantly lower than expected
based on Bolocam observations of other eSZ
cluster candidates which are known to be true clusters.
However, we note the presence of positive flux
immediately south of the SZ decrement for
PLCKESZ G115.71, which may be partially filling
in the SZ decrement from the cluster.
We have also examined WISE 
data at the positions of these cluster candidates,
and find a significant overdensity of sources
towards PLCKESZ G115.71, providing further
evidence that it is a cluster.

\section{Acknowledgments}

We acknowledge the assistance of: 
the day crew and Hilo
staff of the Caltech Submillimeter Observatory, who provided
invaluable assistance during data-taking for this
data set; 
Kathy Deniston, Barbara Wertz, and Diana Bisel, who provided effective
administrative support at Caltech and in Hilo,
and the referees of our manuscript, who provided many useful comments.  
The Bolocam observations were supported by the Gordon and Betty
Moore Foundation.
JS was supported by NSF/AST-0838261
and NASA/NNX11AB07G;
NC was partially supported by a NASA Graduate Student
Research Fellowship;
KU acknowledges support from the Academia Sinica Career
Development Award.
This publication makes use of data products from the 
Wide-field Infrared Survey Explorer, which is a joint
project of the University of California, Los Angeles,
and the Jet Propulsion Laboratory/California 
Institute of Technology, funded by the National
Aeronautics and Space Administration.

\end{document}